\documentstyle[12pt,a4]{article}

\begin{document}

\begin{center}
{\LARGE\bf 
The Drell-Hearn-Gerasimov Sum-Rule } \\
\vspace{3ex}
{\LARGE\bf 
  at Polarized HERA }

\vspace{1cm}
{S. D. Bass $^a$,  M. M. Brisudov{\'a} $^b$ and A. De Roeck $^c$}

\vspace*{1cm}
{\it $^a$
Institut f\"{u}r Theoretische Kernphysik,
Universit\"{a}t Bonn,\\
Nussallee 14--16, D-53115 Bonn, Germany }\\

\vspace*{3mm}
{\it $^b$
Theoretical Division, MSB283, Los Alamos National
Laboratory, Los Alamos, NM 87545, U.S.A. } \\

\vspace*{3mm}
{\it $^c$
DESY, Deutsches Elektronen Synchrotron\\
Notkestrasse 85, D-22607 Hamburg, Germany} \\

\vspace*{1cm}

\end{center}

\begin{abstract}{
\noindent
We discuss the potential of polarized HERA to measure
the spin 
dependent part of the total photoproduction
cross-section at large $\sqrt{s_{\gamma p}}$.}
\end{abstract}

%
\vspace{1mm}
\noindent
The Drell-Hearn-Gerasimov sum-rule~\cite{dhg} 
(for reviews see~\cite{moi, drech})
for spin dependent
photoproduction
relates the difference of the two cross-sections for the absorption
of a real photon with spin anti-parallel $\sigma_A$ and parallel
$\sigma_P$ to the target spin to the square of the anomalous magnetic
moment of the target nucleon,
\begin{equation}
({\rm DHG}) \equiv
- {4 \pi^2 \alpha \kappa^2 \over 2 m^2} =
\int_{\nu_{th}}^{\infty} {d \nu \over \nu} (\sigma_A - \sigma_P)(\nu).
\end{equation}
Here $\nu$ is the (LAB) energy of the exchanged photon,
$m$ is the nucleon mass and $\kappa$ is the anomalous magnetic moment.

The first direct test of the Drell-Hearn-Gerasimov sum-rule will
be made in experiments which are planned or underway at the
ELSA, GRAAL, LEGS and MAMI facilities.
These experiments will measure $(\sigma_A - \sigma_P)$ up to a
photon LAB energy
$\nu = 3$GeV ($\sqrt{s_{\gamma p}} \leq 2.5$GeV).
They will make a precise measurement of nucleon resonance
contributions 
to (DHG),  thus testing multipole analyses of unpolarized 
single-pion photoproduction data~\cite{ikar}, 
as well as the  contributions from strangeness production 
and vector meson dominance.
There is no elastic contribution to (DHG).

The high-energy part of $(\sigma_A - \sigma_P)$ is expected 
to be
determined by Regge theory for $\sqrt{s_{\gamma p}} \geq 2.5$GeV.   
In this note we briefly review possible Regge contributions
to $(\sigma_A - \sigma_P)$ 
and summarise the present knowledge of these contributions 
from low $Q^2$ ($\simeq 0.45$GeV$^2$)
inclusive photoproduction.
Finally, we discuss the contribution that polarized HERA
could make to the measurement of these high $\sqrt{s}$ 
Regge contributions to the Drell-Hearn-Gerasimov sum-rule.

At large centre of mass energy squared ($s = 2m \nu + m^2$),
soft Regge theory predicts~\cite{heim, us, clos1}
\begin{equation}
\Biggl( \sigma_A - \sigma_P \Biggr) \sim
N_3 s^{\alpha_{a_1} - 1} + N_0  s^{\alpha_{f_1} - 1}
+ N_g {\ln {s \over \mu^2} \over s}
+ N_{PP} {1 \over \ln^2 {s \over \mu^2} }
\end{equation}
Here $\alpha_{a_1}$ and $\alpha_{f_1}$ 
are the intercepts of the isovector
$a_1(1260)$ and isoscalar $f_1(1285)$ and $f_1(1420)$ 
Regge trajectories,
which are usually taken between -0.5 and 0.0~\cite{heim}.
(
We note, however, that the isotriplet part of the deep
inelastic structure function $g_1$ 
behaves like 
$x^{-0.5}$ in the range ($0.01 < x < 0.12$) in the SLAC
data~\cite{e154}, 
corresponding to an effective Regge intercept
$\alpha_{a_1}(Q^2) \simeq +0.5$ at 
the relatively low deep inelastic $Q^2 \simeq 3-5$GeV$^2$.) 

The $\ln s / s$ term is induced by any vector component
to the short range exchange potential~\cite{clos1}
-- for example, 
two nonperturbative gluon exchange~\cite{sbpl}.
The $1 / \ln^2 s$ term represents any pomeron-pomeron cut
contribution to $(\sigma_A - \sigma_P)$.
The mass parameter $\mu$ is a typical hadronic $\sim 0.5$GeV.
The coefficients $N_3$, $N_0$, $N_g$ and $N_{PP}$ 
in  Equ.(2) are to be determined from experiment.

Besides their importance for a precise test of the DHG
sum-rule,
the soft Regge contributions to $(\sigma_A - \sigma_P)$
form a baseline for investigations of DGLAP and BFKL
small $x$ behaviour in $g_1$, the nucleon's first spin
dependent structure function.

To estimate the Regge contribution to $(\sigma_A - \sigma_P)$ 
at $Q^2=0$  
we take the low $Q^2$ data from the E-143~\cite{e143} and 
SMC~\cite{smc} experiments
(0.25GeV$^2 < Q^2 < 0.7$GeV$^2$) on
\begin{equation}
A_1 = {\sigma_A - \sigma_P \over \sigma_A + \sigma_P}
\end{equation}
This low $Q^2$ data has the following features.
First,
the spin asymmetries $A_1^p$ and $A_1^d$
show no significant
$Q^2$ dependence in the range of each experiment.
Secondly,
the isoscalar deuteron asymmetry $A_1^d$
is consistent with zero in both the E-143 and SMC
low $Q^2$ bins.
There is a clear positive proton asymmetry
in the
E-143 data,
signalling a strong isotriplet term
in $(\sigma_A - \sigma_P)$ at $s \simeq 12$GeV$^2$.
The SMC $A_1^p$ data is less clear:
combining the SMC low $Q^2$ $A_1^p$ data yields a 
positive value for $A_1^p$.
However, the majority of these SMC points are consistent with zero.

In Table 1 we combine the low $Q^2$ data to obtain 
one point corresponding to each experiment.
We impose the cut
($Q^2 \leq 0.7$GeV$^2$, $\sqrt{s} \geq 2.5$GeV),
so that the mean $Q^2$ is less than $0.5$GeV$^2$ for each experiments,
and so that
our data set is well 
beyond the
resonance region.
\begin{table}
\begin{center}
\caption{$A_1$ at large $s$ and low $Q^2$}
\begin{tabular} {ccccc}
\\
\hline\hline
\\
Experiment & $\langle Q^2 \rangle$ & $\sqrt{s}$ & $A_1^p$ & $A_1^d$ \\
\\
\hline
\\
($Q^2 \leq 0.7$GeV$^2$, $\sqrt{s} \geq 2.5$GeV) & & & & \\
\hline
E-143  &  0.45  &  3.5  &  $ 0.077 \pm 0.016 $  & $ +0.008 \pm 0.022 $ \\
SMC    &  0.45  &  16.7 &  $ 0.064 \pm 0.024 $  & $ -0.013 \pm 0.020 $ \\
\hline\hline
\end{tabular}
\end{center}
\end{table}

We assume that the large $\sqrt{s}$ $A_1$ is approximately
independent of $Q^2$
between $Q^2=0$ and $Q^2 \simeq$0.5 GeV$^2$.
For the total photoproduction cross-section we take
\begin{equation}
(\sigma_A + \sigma_P) = 67.7 s^{+0.0808} + 129 s^{-0.4545}
\label{sigtot}
\end{equation}
(in units of $\mu$b),
which is known to provide a good Regge fit for $\sqrt{s}$ 
between 2.5GeV and 250GeV~\cite{pvl1}.
(Here, 
the $s^{+0.0808}$ contribution is associated with pomeron
exchange
and the $s^{-0.4545}$ contribution is associated with the 
isoscalar $\omega$ and isovector $\rho$ trajectories.)
Since the E-143 low $Q^2$ data shows a clearly positive $A_1^p$
with the smallest experimental error, we choose to normalise to
E-143.
We estimate
\begin{equation}
(\sigma_A - \sigma_P) \simeq 10 \mu{\rm b} \ \ \ \ 
{\rm at} \ \ \ \ 
(Q^2=0, \sqrt{s} = 3.5 {\rm GeV})
\end{equation}

The small isoscalar deuteron asymmetry $A_1^d$ indicates that
the isoscalar contribution to $A_1^p$ in the E-143 data is unlikely 
to be more than 30\%.
In Fig.~1 we show the asymmetry $A_1^p$ as a function of $\sqrt{s}$ 
between 2.5 and 250 GeV
for the 
four different would-be Regge behaviours for
$(\sigma_A - \sigma_P)$:
that the high energy behaviour of $(\sigma_A-\sigma_P)$ is given
\begin{enumerate}
\item
entirely by the $(a_1, f_1)$ terms in Equ.(2) with Regge 
    intercept either (a)
    $-{1 \over 2}$ (conventional) or
    (b)  $+{1 \over 2}$
    (motivated by the observed small $x$ behaviour of $g_1^{(p-n)}$),
\item
by taking 2/3 isovector (conventional) $a_1$ and 
1/3 two non-perturbative 
gluon exchange contributions at $\sqrt{s} = 3.5$GeV,
\item
by taking 2/3 isovector (conventional) $a_1$ and 1/3 
pomeron-pomeron cut
contributions at $\sqrt{s} = 3.5$GeV.
\end{enumerate}

%

Photoproduction cross-sections can be measured in $ep$ 
collisions at HERA at high 
$\sqrt{s_{\gamma p}}$ energies. The dominant processes in 
$ep$ collisions are $\gamma p  $ interactions where the photon is on
mass shell. The electron is scattered under approximately zero degrees
with respect to the electron beam direction, 
which at HERA means that  it remains
in the beampipe. The energy of the scattered electron 
$E_e'$ is however reduced to $E_e' = E_e -E_{\gamma}$, with $E_e$
the incident electron  energy and $E_{\gamma}$  the emitted photon energy.
The HERA machine magnets in the beamline, 
which steer the beam into a closed orbit, 
act as a spectrometer on these off-momentum electrons. 
The experiments H1 and ZEUS  
have installed calorimeters to detected these kicked out electrons along
the beamline.
In case of H1 calorimeters (stations) 
are installed at three locations: at 8 m , 30 m  and 44 m
distance from the interaction point~\cite{H1}.
 The stations accept (tag) electrons from 
different momentum ranges, which correspond to $\sqrt{s_{\gamma p}}$ ranges 
of 280-290 GeV, 150-250 GeV and 60-115 GeV respectively.
At the central energy value of each region the acceptance of these devices 
amounts to 15\%, 85\% and 70\% respectively.
The locations of the stations may change as a result of the HERA 
luminosity upgrade, and are presently under study.

Equ.\ref{sigtot} shows that the $\gamma p$ cross section is large, of order
of hundreds of microbarns. The photon energy spectrum 
emitted from an electron beam is given in the 
 Weizs\"acker-Williams approximation~\cite{www}. 
An integrated $ep$ luminosity 
at HERA of 1 pb$^{-1}$ can yield about $N= 1500$K $ \gamma p$ events
in each of the 30 m and 44 m stations, and about 10 times less in the 8 m
station. 

The collider experiments record presently unpolarized $ep$ 
collisions. Around the year 2000, spin rotators will be installed 
for the electrons, converting the natural transverse polarization 
of the electron beam into  a physics whise more useful longitudinal one. 
Studies are
being made to have also the proton beam at HERA polarize~\cite{barber}
which would enable polarized $ep$ and thus polarized $\gamma p$ 
collisions. Note however that the polarization of the photon beam 
will be reduced by a factor $D = y(2-y)/(y^2+2(1-y))$, the so called
depolarization factor.
Here $y = s_{\gamma p}/s_{ep}$. For the three stations the measurements
are at $y= 0.09, 0.44$ and 0.90, leading to values of $ D = 0.094, 0.52$
and 0.98. The measured asymmetries  at HERA are correspondingly reduced 
by this factor.

If HERA is fully polarized, event samples of the order of 100 - 500 
pb$^{-1}$ will be collected, with expected beam polarizations 
$P_e =  P_p =  0.7 $. In the small asymmetry approximation the 
error on the asymmetries, $\delta A_1$ can be calculated as
$1/(P_eP_p\sqrt{N})$.  Due to data-taking bandwidths and 
trigger problems presently not all tagged $\gamma p$ events 
are recorded. Assuming a data taking rate of 2 Hz for these events,
 also in 
future, leads to  about 40 M events/year
giving  a reachable precision 
  $\delta A_1 = 1/(P_eP_p\sqrt{N})= 0.0003$.
It is however not excluded that novel techniques in 
triggering, data-taking and data storage will become available and can be 
installed for the experiments, which would allow to
collect all produced events, amounting to approximately
15,000M events in total for  the 
three stations in a period of 3 to 5 years. This would lead to  
 maximal reachable 
precisions of $\delta A_1 = 3.10^{-5} $
for a measurement at the 30m and the 44m station, and $\delta A_1 = 10^{-4} $
for a measurement at the 8m station.
Note that when compared with the true asymmetries as shown in Fig.~1,
the depolarization factor $D$ reduces the
effective  sensitivities  with the 
numbers given above.

Given the projected asymmetries, polarized HERA 
with $\delta A_1 = 0.0003$ would be sensitive 
at $\sqrt{s_{\gamma p}}=50$GeV to a 
$(\sigma_A-\sigma_P)$ falling no faster than $s^{-1}$.
Taking the conventional,
$\alpha_{a_1} = - {1\over 2}$,
it would be sensitive to a two non-perturbative gluon exchange 
contribution which is not less than 30\% in the E-143 data and 
to a pomeron-pomeron cut
contribution which is not less than 3\% in the E-143 data.
At
$\sqrt{s_{\gamma p}}=250$GeV, 
polarized HERA would be sensitive to
$(\sigma_A-\sigma_P)$ which falls no faster than $s^{-0.6}$
and to a pomeron-pomeron 
cut contribution which is no less than 6\% of 
the E-143 $A_1^p$.
We note that a zero-result (no significant signal) 
would put an upper bound on the Regge contribution 
to the Drell-Hearn-Gerasimov integral.

\vspace{1.0cm}

Support from the Alexander von Humboldt Foundation (SDB) 
and the United States Department of Energy (MMB) 
is gratefully acknowledged.

\newpage

\begin{figure}
\centerline{}
\begin{center}
\input{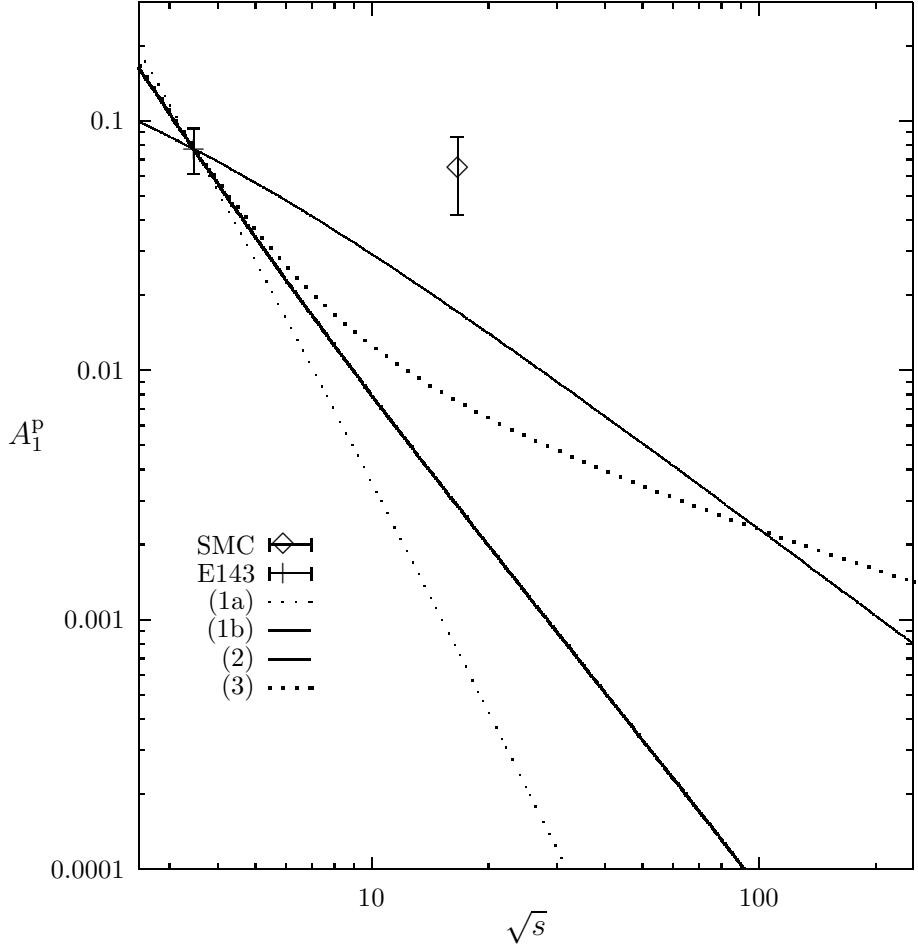}
\caption{
{\bf The asymmetry $A_1^p$ as a function of $\sqrt{s}$ 
for  different  Regge behaviours } for
$(\sigma_A - \sigma_P)$:
 given
entirely by (1a) the $(a_1, f_1)$ terms in Equ.(2) with Regge 
    intercept either 
    $-{1 \over 2}$ (conventional) or
    (1b)  $+{1 \over 2}$;
(2) by  2/3 isovector (conventional) $a_1$ and 
1/3 two non-perturbative 
gluon exchange contributions at $\sqrt{s} = 3.5$GeV;
(3)
by 2/3 isovector (conventional) $a_1$ and 1/3 
pomeron-pomeron cut
contributions at $\sqrt{s} = 3.5$GeV.
}
\end{center}
\end{figure}

\end{document}